\documentclass[twocolumn,amsmath,amssymb]{revtex4}
\usepackage[english]{babel}
\usepackage{graphicx}

\begin{document}

\title{Third-order Self-action Effects in Photonic Microcavities}

\author{I. Razdolski, R. Kapra, T. Murzina, O. Aktsipetrov}

\affiliation{Physics Department, Moscow State University, 119991 Moscow,
Russia}

\author{M.\,Inoue}

\affiliation{Toyohashi University of Technology, 441-8580, Toyohashi, Japan\\}

\begin{abstract}Third-order nonlinear optical effects in photonic microcavities are studied. Significant light defocusing in the thin nonlinear microcavity spacer was observed. The polarization self-action effect was detected, when the large nonlinear polarization rotation angle arises when exciting the microcavity mode, being proportional to the radiation intensity.
\end{abstract}

\maketitle

Photonic crystals and microcavities (MC), structures with the periodic modulation of the refraction index, possessing the photonic bandgap (PBG), became recently object of the intensive studies
\cite{yablonovitch}, \cite{1dmpmc}, \cite{fainstein}.  MCs with the microcavity mode inside the PBG provide conditions for the light to propagate inside the MC without the decrement, when the mode is excited. MCs with the nonlinear MC spacer are promising objects for the nonlinear optics, while one can expect the significant enhancement of the nonlinear-optical effects due to the light localization and its multipassing propagation. Light localization in the MC spacer leads to the optical field enhancement, while the multipassing can be described in terms of effective increase of the thickness of the active layer. The enhancement of the parametric nonlinear-optical effects was observed for the second \cite{pellegrini}, \cite{shgmpc} and third \cite{eleven} harmonic generation.

In this paper the nonparametric third-order nonlinear-optical effects in MCs are studied, namely, self-action effects, which relate to the refractive index dependence on the light intensity. The significant self-defocusing of light and self-induced polarization rotation were observed in the nonlinear dielectric layer, its thickness being much smaller than the radiation wavelength.

The samples studied were sputtered on the fused silica substrate, starting from the one-dimensional photonic crystal of 5 pairs of (SiO$_2$/Ta$_2$O$_5$) layers, optical thickness of each layer corresponding to the $\lambda /4$, where $\lambda$ is the wavelength of the microcavity mode at normal incidence. Then the $\lambda/2$ MC spacer of polycrystalline bismuth-doped yttrium-iron garnet was sputtered, as thick as 225 nm, which corresponds to the $\lambda\simeq 890$ nm. The second photonic crystal mirror, identical to the first one, was deposited at the top of the structure.

 \begin{figure}
  \begin{centering}
  \includegraphics[width=\columnwidth]{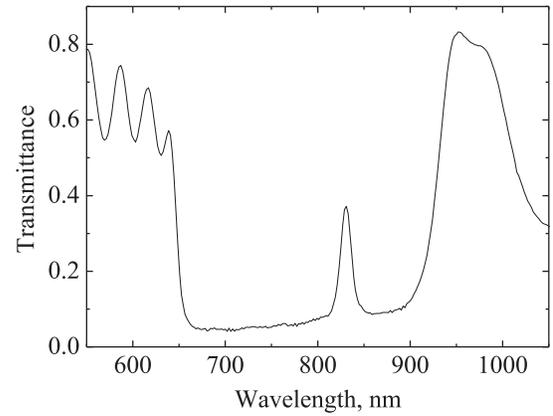}
   \caption{Transmittance spectrum of the MC studied, angle of incidence about 45$^0$.}
  \end{centering}
\end{figure}

Fig. 1 represents the MC transmission spectrum at the 45$^0$ of incidence. Low transmission in the region of 650-900 nm ($\textless$ 0.1) is attributed to the PBG of the MC mirrors, while the narrow peak at the $\simeq$ 830 nm corresponds to the MC mode with the quality factor $Q_{MP}\simeq 75$ \cite{msthg}.

When studying the third-order nonlinear effects the sample was excited by the radiation of the femtosecond Ti:Sapphire oscillator with the pulsewidth of 80 fs, repetition rate of 82 MHz, average power of 150 mW and wavelength of 830 nm. Angle of incidence was about 45$^0$, in order to excite the microcavity mode.

For the third-order effect study the closed-aperture z-scan method was used, first suggested in \cite{sheik-bahae}. The scheme of the experimental setup used is shown at Fig. 2,a. Laser radiation passed through the Glan prism was focused on the sample by the lens with F = 6 cm, while the sample was translated along the beam direction by means of the step motor in the vicinity of the lens focal plane, effectively changing the power density of the laser radiation exciting the sample.

When studying self-action the radiation transmitted through the MC was cut off with the diaphragm and the red filter and then registered by the photodiode. We measured the normalized transmittance $T$ against the sample position with respect to the lens focal plane z, where $z = 0$ corresponds to the focal plane, and $T = 1$ denoted the transmission far away from $z =0$.

Fig. 2,b shows the $T(z)$ dependence obtained. The light defocusing was therefore observed, being determined by the intensity-dependent refractive index as follows: $n(I)=n{_0} + n_2^{eff} I = n{_0} + n{_2} Q^{2} I$, where $n_0$ is the refractive index of the nonlinear layer, and $n_2^{eff}$ denotes the effective nonlinear refraction coefficient. The $n_2^{eff}$ value is determined by the real part of third-order susceptibility $\chi^{(3)} = \frac{n_0 n_2}{3\pi}$ and the MC quality factor $Q$. The latter leads to the intensity of light increase in the nonlinear MC spacer due to the the multibeam interference and multipassing propagation. The effective nonlinear refractive coefficient obtained was $n_2^{eff} = -(3.9\pm 0.6)\cdot 10^{-9}$ cm$^2$/W, being much larger than the typical value. The $n_2^{eff} / n_2$ ratio can be as large as four orders of magnitude in our samples.

To study the polarization rotation dependence on the light intensity the polarization-sensitive open-aperture z-scan modification was developed. In this scheme (see Fig. 2,c) the normalizing dependence $T(z)$ should first be measured, referred below as $I_0(z)$. Open-aperture z-scan here means that no diaphragm is needed any more in contrast with the closed-aperture z-scan method described above. Further, the transmitted radiation was partially cut-off by another Glan prism with its axis directed at the $+45^0$ or $-45^0$ with respect to the first one (see scheme), and the dependences $I_+(z)$, $I_-(z)$ were obtained. Finally, the polarization rotation angle was determined using the following equation:

\begin{equation}\label{method}
\sin 2\theta(z)=\frac{I_{-}(z)-I_{+}(z)}{I_{0}(z)},
\end{equation}

\begin{figure}
\begin{centering}
\includegraphics[width=\columnwidth]{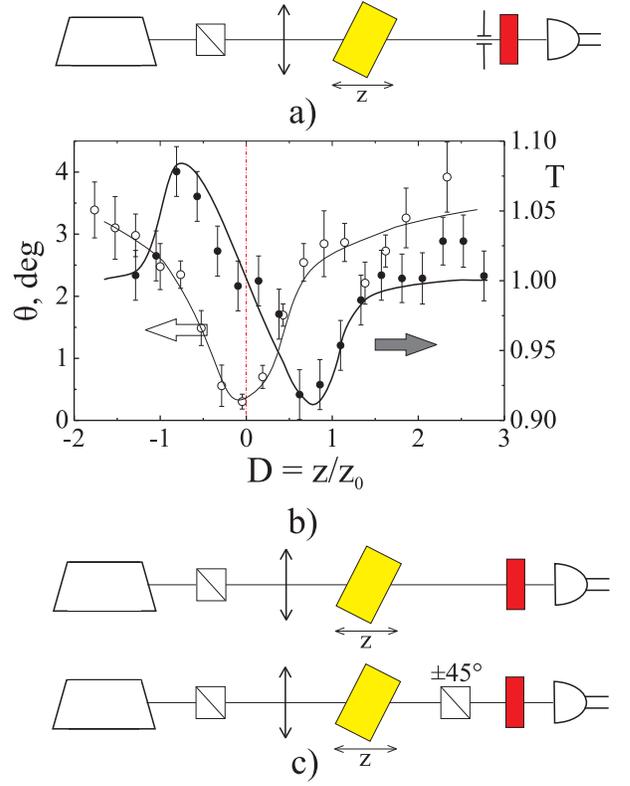}
\caption{(a) Closed-aperture z-scan scheme. (b)
Normalized transmittance $T$ in closed-aperture z-scan (filled) and polarization rotation angle $\theta$ (empty) versus the sample position $z$. $z_0$ is the beam diffraction length of about 0.7 cm. (c) Polarization-sensitive open-aperture z-scan scheme.}
  \end{centering}
\end{figure}

In the similar way the linear polarization rotation angle spectrum was obtained, when the light intensity is low and all the nonlinearities are negligible. The bulb was used as a radiation source, and the wavelength $\lambda$ substitutes the sample position $z$ in (\ref{method}). The spectrum obtained is presented at the inset of the Fig. 3. For the radiation exciting the microcavity mode the linear polarization rotation angle was found to be $(4.3\pm 0.3)^0$. The rotation can be attributed to the birefringence of the microcavity spacer. Fig. 3 shows the dependence of the nonlinear contribution into the total rotation angle $|\Delta\theta (\lambda)|=|\theta (\lambda)-\theta_0|$ on the radiation intensity. The clear linear dependence obtained allows to conclude that the effect observed is third-order.

\begin{figure}
  \begin{centering}
   \includegraphics[width=\columnwidth]{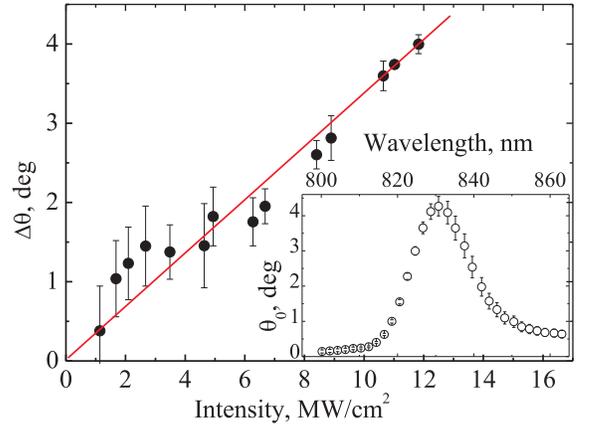}
   \caption{Nonlinear contribution into the polarization rotation angle $|\Delta\theta|$ versus the radiation intensity. Inset: linear polarization rotation angle spectrum in the vicinity of the MC mode at 45$^0$ angle of incidence.}
  \end{centering}
\end{figure}

The polarization rotation can be driven by the birefringence expected in the MC spacer. Following \cite{msthg}, the sample preparation technique leads to the refractive index anisotropy along the normal direction with respect to the layer plane, so the MC spacer becomes single-axis birefringent. In this case the birefringence should be function of the angle of incidence. The transmittance spectra of the MC were measured at different angles of incidence for two orthogonal beam polarizations. Inset at the Fig. 4 shows the MC mode spectral position dependence on the angle of incidence; in the Fig. 4 the effective birefringence $\Delta n=n_p-n_s$ angular spectrum is presented, where $n_p$, $n_s$ are the refraction indices for the $p-$ and $s-$polarized radiation.

According to \cite{shen}, birefringence leads to the polarization plane rotation, with the angle of rotation being proportional to the birefringence value. When using the same formula for the $n_p$, $n_s$ indices, one can obtain:

\begin{equation}\label{nlbiref}
\theta\propto\Delta n = (n_{s0} - n_{p0})+(n_{s2} - n_{p2})I
\end{equation}

\begin{figure}[!t]
  \begin{centering}
   \includegraphics[width=\columnwidth]{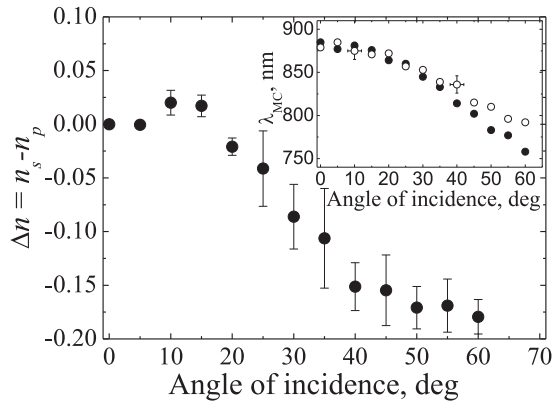}
   \caption{The birefringence $\Delta n=n_s - n_p$ dependence on the
angle of incidence. Inset: Spectral MC mode position $\lambda_{MC}$ dependence   on the angle of incidence for the $s-$ (empty) and $p-$ (filled) polarized radiation.}
  \end{centering}
\end{figure}

The large effect magnitude can also be attributed to the MC nature of the sample, as well as the light self-defocusing. Let us consider the homogeneous film of bismuth-doped YIG the same thickness as the MC spacer in the sample studied; if the linear polarization rotation angle is $\theta_0$, in the nonlinear case the intensity dependence arises, which can be described by $\theta =\theta_0 + \theta_2 I_0$, where $I_0$ is the light intensity inside the film and also the laser beam intensity, and $\theta_2$ represents the third-order contribution. In MC structure one can expect that multipassing propagation will enhance the effect magnitude by Q times. Finally, one should obtain $\theta_{MC}=Q(\theta_0 + \theta_2 I_{MC})= Q(\theta_0 + Q\theta_2 I)$, due to the field intensity inside the MC also exhibits enhancement. So the third-order contribution enhancement factor appears to be about $Q^2$, or four orders of magnitude, as well as $n_2$ effective contribution.

In conclude, the light defocusing was observed in the nonlinear microcavity; the effective nonlinear contribution into the refractive index at 830 nm wavelength was found to be $n_2^{eff} = -(3.9\pm 0.6)\cdot 10^{-9}$ cm$^2$/W, being much larger than the typical $n_2$ value. The nonlinear contribution into the polarization rotation angle was also observed in the vicinity of the MC mode. The polarization effect was shown to origin from the birefringence of the MC spacer.

We would like to note that the MC layer in our sample is made of ferromagnetic material, so one can also expect the magnetic contribution into the self-action of light or the polarization rotation (nonlinear Faraday rotation, \cite{Granovsky}). In this case the non-diagonal dielectric permeability component should have the nonlinear (intensity-dependent) contribution as well.

The work is supported by the RFBR Grants 04-02-16487, 04-02-17059, 05-02-19886, 06-02-91201.

\vfill\eject


\begin{thebibliography}{99}

\bibitem{yablonovitch} E. Yablonovitch, Phys. Rev. Lett.,
\textbf{58}, 2059(1987).

\bibitem{1dmpmc} M. Inoue, K. Arai, T. Fujii, M. Abe, J. Appl.
Phys., \textbf{85}, 5768 (1999)

\bibitem{fainstein} A. Fainstein, B. Jusserand, V. Thierry-Mieg,
Phys.Rev.Lett., \textbf{74}, 3764(2003).

\bibitem{pellegrini} V. Pellegrini, R. Colombelli, I. Carusotto et
al., Appl. Phys. Lett. \textbf{74}, 1945 (1999).

\bibitem{shgmpc} T.V. Dolgova, A.I. Maidykovsky, M.G. Martemyanov,
A.A. Fedyanin, G. Marowsky, V.A. Yakovlev, G. Mattei, Appl. Phys.
Lett., \textbf{81} 2725 (2002)

\bibitem{eleven} T.V. Dolgova, A.I. Maidykovsky, M.G. Martemyanov,
A.A. Fedyanin, O.A. Aktsipetrov, JETP Letters, \textbf{75}, 17
(2002) (in russian).

\bibitem{msthg} O.A. Aktsipetrov, T.V. Dolgova, A.A. Fedyanin, T.V.
Murzina, M. Inoue, K. Nishimura, H. Uchida, J. Opt. Soc. Am. B
\textbf{22}, 176 {2005}.

\bibitem{sheik-bahae} M. Sheik-Bahae, A. A. Said, E. W. Van Stryland,
Optics Letters, \textbf{14}, 955 (1989).

\bibitem{shen} I.R. Shen, "The Principles of Nonlinear Optics", John Wiley and Sons, 1984.

\bibitem{Granovsky} A. B. Granovsky, M. V. Kuzmichov, J.-P. Clerc,
M. Inoue, JMMM, \textbf{258} 103 (2003).

\end{thebibliography}
\end{document}